\documentstyle[12pt]{article}
\makeatletter
\def\appendix{\par\clearpage
  \setcounter{section}{0}
  \setcounter{subsection}{0}
  \@addtoreset{equation}{section}
  \def\@sectname{Appendix~}
  \def\theequation{\thesection.\arabic{equation}}
  \def\thesection{\Alph{section}}}
\makeatother

\renewcommand{\theequation}{\thesection.\arabic{equation}}

\begin{document}

\begin{titlepage}
\hskip 11cm \vbox{ \hbox{Budker INP 2002-38} \hbox{DESY  02-074}
\hbox{DFCAL-TH 02/2} \hbox{June 2002}} \vskip 0.3cm
\centerline{\bf A PROOF OF FULFILLMENT OF THE STRONG BOOTSTRAP
CONDITION$^{~\ast}$} \vskip 0.8cm
\centerline{V.S.~Fadin$^{a~\dagger}$ and A.~Papa$^{b~\ddagger}$}
\vskip .3cm \centerline{\sl $^a$ Budker Institute for Nuclear
Physics, 630090 Novosibirsk, Russia} \centerline{\sl and
Novosibirsk State University, 630090 Novosibirsk, Russia}
\centerline{\sl $^b$ Dipartimento di Fisica, Universit\`a della
Calabria,} \centerline{\sl and Istituto Nazionale di Fisica
Nucleare, Gruppo collegato di Cosenza,} \centerline{\sl I-87036
Arcavacata di Rende, Cosenza, Italy} \vskip 1cm

\begin{abstract}
It is shown that the kernel of the BFKL equation for the octet color
state of two Reggeized gluons satisfies the strong bootstrap condition in
the next-to-leading order. This condition is much more restrictive than the one
obtained from the requirement of Reggeized form for the elastic scattering
amplitudes in the next-to-leading approximation.
It is necessary, however, for self-consistency of the assumption of Reggeized
form of the production amplitudes in multi-Regge kinematics, which are used
in the derivation of the BFKL equation. The fulfillment of the strong bootstrap
condition for the kernel opens a way to a rigorous proof of the BFKL equation
in the next-to-leading approximation.
\end{abstract}
\vfill

\hrule \vskip.3cm

\noindent $^{\ast}${\it Work supported in part by the Ministero
dell'Istruzione, dell'Universit\`a e della Ricerca, in part by INTAS and 
in part by the Russian Fund of Basic Researches.} 
\vfill $ \begin{array}{ll} ^{\dagger}\mbox{{\it
e-mail address:}} &
\mbox{FADIN@INP.NSK.SU}\\
^{\ddagger}\mbox{{\it e-mail address:}} &
\mbox{PAPA@FIS.UNICAL.IT}\\
\end{array}
$
\vfill

\end{titlepage}
\eject

\section{Introduction}
\setcounter{equation}{0}

The BFKL approach~\cite{BFKL} to the description of processes at large c.m.s.
energy $\sqrt s$ and fixed momentum transfer $\sqrt -t$ is based on the gluon
Reggeization. In this approach the scattering amplitudes are given by the
convolution of the impact factors of the scattered particles and the Green
function for the Reggeon-Reggeon scattering.
The Pomeron, which determines the high energy behavior of cross sections, and
the Odderon, responsible for the difference of particle and antiparticle cross
sections, appear in this approach as compound states of two and three Reggeized
gluons respectively.

Originally the BFKL approach was developed in the leading logarithmic
approximation (LLA), that means resummation of terms $(\alpha_s \ln s)^n$.
Now for the forward scattering ($t=0$ and color singlet in the $t$-channel) the
kernel of the equation for the two-Reggeon Green function is known~\cite{FL98,CC98}
in the next-to-leading order (NLO). The calculation of the NLO kernel for the
non-forward scattering~\cite{FF98} is not far from completion
(see~\cite{FFP99,FG}). The impact factors of gluons~\cite{FFKPg} and
quarks~\cite{FFKPq} are calculated in the NLO and the impact factors of the
physical (color singlet) particles are under
investigation~\cite{FM99,FIK1,BGQ,FIK2,BGK}.

Note that although initially the gluon Reggeization appeared as the assumption
used in the derivation of the BFKL equation~\cite{BFKL}, later it was proved
in the LLA~\cite{BLF}. The key to the proof is given by the properties of the
amplitudes with gluon quantum numbers (color octet with negative signature)
in the $t$-channels.

Whereas two Reggeized gluons in the color singlet state create the BFKL Pomeron,
in the antisymmetric color octet state they must reproduce the Reggeon itself.
This is the requirement of self-consistency of the BFKL approach, since it is
based on the gluon Reggeization. The bootstrap conditions for the color octet
impact factors and the BFKL kernel in the NLO, arising from the application of
this requirement to the elastic scattering amplitudes, were obtained
in~\cite{FF98}. The condition for the impact factors was checked
in~\cite{FFKPg} and~\cite{FFKPq} in the cases of gluon and quark, respectively,
and was proved to be satisfied at arbitrary space-time dimension $D$
both for the helicity conserving and non-conserving parts of the impact factors,
as well as the bootstrap condition for the quark part of the kernel, which was
analyzed in~\cite{FFP99}. The fulfillment of the bootstrap condition for the gluon
part of the kernel was proved in the limit $D\rightarrow 4$~\cite{FFK00}.
The evident reason for this restriction was that the gluon part of the color
octet kernel was known at that time~\cite{FG} only in this limit.
Note that now the kernel at arbitrary $D$ is calculated~\cite{FFP00} and it is
possible to check the bootstrap condition at arbitrary $D$.

Although the bootstrap conditions arising from the elastic scattering amplitudes
are rather restrictive, their fulfillment can not justify completely the BFKL
equation. Evidently, the bootstrap conditions must be satisfied for all amplitudes
involved into the derivation of the BFKL equation, since they were supposed to
have (multi) Regge form, i.e. for the amplitudes of production of any number of
particles in the multi-Regge and (in the NLA) quasi multi-Regge kinematics. The
fulfillment of all these conditions gives the basis on which the proof of
Reggeization was constructed in the LLA~\cite{BLF}. An analogous
proof can be constructed in the next-to-leading approximation (NLA) as
well~\cite{tbp}.

The requirement of bootstrap for the multi-particle production leads, in 
particular,~\cite{tbp} to stronger restrictions on octet impact factors and kernel, 
than it follows from the elastic amplitudes. These restrictions are just the so 
called strong bootstrap conditions, which were suggested, without derivation,
in~\cite{B99,BV00}. They can be presented in the form~\cite{FFKP00}:
\begin{equation}
\Phi^{a}_{A^\prime A}(\vec q_1,\vec q) =\frac{-ig\sqrt{N}}{2}\Gamma_{A^\prime
A}^{a}(q)R(\vec q_1,\vec q)~,
\label{boot2}
\end{equation}
\begin{equation}
\int\frac{d^{D-2}q_2}{\vec q_2^{\:2}\vec q_2^{\:\prime \:2}}{\cal K}(\vec q_1,
\vec q_2;\vec q)R(\vec q_2,\vec q) =\omega(t)R(\vec q_1,\vec q)~.
\label{boot1}
\end{equation}
Here and in the following $N_c$ is the number of colors, $\vec q$ is the
transverse part of the total transferred momentum, $t=-\vec q^{\:2}$,
$\vec q_i$ and $\vec q_i^{\:\prime}\equiv \vec q-\vec q_i$, $i=1,2$, are the momenta
of the Reggeized gluons; $\Phi^{a}_{A^\prime A}$ is the octet impact factor for
the $A\rightarrow A^\prime$ transition, $\Gamma_{A^\prime A}^{a}$ is the Reggeon
vertex, $R(\vec q_1,\vec q)$ appears as an universal (i.e. independent on
properties of the scattered particles) function, which occurs to be the
eigenfunction of the octet kernel ${\cal K}(\vec q_1, \vec q_2;\vec q)$ with
eigenvalue $\omega(t)=j(t)-1$, $j(t)$ is the gluon Regge trajectory. Note that
here we have changed the normalization of $R$ compared to~\cite{FFKP00}, so that
now in the leading order $R=R^{(0)}=1$. Fulfillment of~(\ref{boot2}) in this order
is immediately seen from explicit expressions for the impact factors and
vertices, which are constant in this order. The remarkable property of the
color octet impact factors and Reggeon vertices exhibited by this relation is
that their ratio is a process-independent function. This property becomes
quite nontrivial in the NLO. It was verified by comparison of this
ratio for quarks and gluons in~\cite{FFKP00}, where the NLO contribution
$R^{(1)}$ to the function $R$ was determined.

Taking into account a general relation~\cite{FF98}:
\begin{equation}
{\cal K}(\vec q_1,\vec q_2;\vec q)=\biggl(\omega(-\vec q_1^{\:2})+
\omega(-\vec q_1^{\:\prime\:2})\biggr)\:\vec q_1^{\:2}\:\vec q_1^{\:\prime\:2}
\delta^{(D-2)}(\vec q_1-\vec q_2)+{\cal K}_r(\vec q_1,\vec q_2; \vec q)\;,
\label{BFKL}
\end{equation}
where ${\cal K}_r$ is the part of the kernel related to the real particle
production, the condition~(\ref{boot1}) can be rewritten as
\begin{equation}
\int\frac{d^{D-2}q_2}{\vec q_2^{\:2}\vec q_2^{\:\prime \:2}}{\cal K}_r(\vec q_1,
\vec q_2;\vec q)R(\vec q_2,\vec q) =\left(\omega(t)-\omega(t_1)-\omega(t_1^\prime)
\right)R(\vec q_1,\vec q)~,
\label{boot}
\end{equation}
where $t_i=q_i^2=-\vec q_i^{\:2}$, $i=1,2$.
The fulfillment of this relation in the leading order (remember that $R^{(0)}=1$):
\begin{equation}
\int\frac{d^{D-2}q_2}{\vec q_2^{\:2}\vec q_2^{\:\prime \:2}}{\cal K}^{(0)}_r
(\vec q_1, \vec q_2;\vec q)=\omega^{(1)}(t)-\omega^{(1)}(t_1)-\omega^{(1)}
(t_1^\prime)
\label{boot0}
\end{equation}
is evident, taking into account the leading order expressions for the ``real''
part of the kernel
\begin{equation}
{\cal K}_r^{(0)}(\vec q_1,\vec q_2; \vec q) = \frac{g^2 N_c}{2(2\pi)^{D-1}}\:
f_B(\vec q_1,\vec q_2; \vec q) \;,
\;\;\;\;\;\;\;\;\;\;
f_B(\vec q_1,\vec q_2; \vec q) = \frac{\vec q_1^{\:2} \vec q_2^{\:\prime\:2}
+\vec q_2^{\:2} \vec q_1^{\:\prime\:2}}{\vec k^{\:2}}-\vec q^{\:2}
\;,
\label{K_Born}
\end{equation}
where $\vec k =\vec q_1 - \vec q_2$, and the trajectory
\begin{equation}
\omega^{(1)}(t) =\frac{g^2N_c t}{2(2\pi)^{D-1}} \int\frac{d^{D-2}q_1}{\vec
q_1^{\:2}\vec q_1^{\:\prime \:2}}=-g^2 \frac{N_c \Gamma(1-\epsilon)}{(4 \pi)^{D/2}}
\frac{\Gamma^2(\epsilon)}{\Gamma(2\epsilon)} (\vec q^{\:2})^\epsilon \;,
\label{omega_1}
\end{equation}
with $D=4+2\epsilon$.

In this paper we prove the fulfillment of the bootstrap condition~(\ref{boot})
in the NLO. It is explicitly formulated and
simplified in the next Section; Section 3 contains the proof. The
details of the calculations are given in the Appendices. The
significance of the result is discussed in Section 4.

\section{Formulation and simplification of the bootstrap condition}

Let us work in pure gluodynamics, since the fulfillment of the strong
bootstrap condition for the quark part of the kernel simply follows from
the results of~\cite{FFP99,BV00,BV99}. Then the condition~(\ref{boot2}),
with account of $R^{(0)}=1$ and~(\ref{boot0}) gives us in the NLO
\[
\int\frac{d^{D-2}q_2}{\vec q_2^{\:2} \vec q_2^{\:\prime\:2}}
\biggl[{\cal K}_r^{(1)}(\vec q_1,\vec q_2, \vec q)+
{\cal K}_r^{(0)}(\vec q_1,\vec q_2, \vec q)\: \left(R^{(1)}(\vec q_2, \vec q)-
 R^{(1)}(\vec q_1, \vec q)\right)
\]
\begin{equation}
= \omega^{(2)}(t) -\omega^{(2)}(t_1)-\omega^{(2)}(t_1^\prime)~.
\label{bootstrap}
\end{equation}
Here
${\cal K}_r^{(1)}$, $\omega^{(2)}$ and $R^{(1)}$ are the NLO contributions to
the ``real'' part of the kernel ${\cal K}$ of the BFKL equation for the octet
color representation in the $t$-channel, to the gluon Regge trajectory and to
the function
$R$ determined by the bootstrap condition for the impact factors~(\ref{boot2}).
We have
\begin{equation}
\omega^{(2)}(t) = \left[\frac{g^2N_c \Gamma(1-\epsilon)(\vec q^{\:2})^\epsilon}
{(4 \pi)^{D/2}\epsilon}\right]^2
\left[\frac{11}{3}+\left(2 \psi'(1)-\frac{67}{9}\right)\epsilon
+\left(\frac{404}{27}+\psi''(1)-\frac{22}{3}\psi'(1)\right)\epsilon^2\right]
\;,
\label{omega_2}
\end{equation}
where $\psi (z)= \Gamma^{\prime}(z)/\Gamma (z)$ and all non-vanishing terms for
$\epsilon\rightarrow 0$ are held. Although the expression of
$\omega^{(2)}(t)$ in terms of integrals in the transverse momentum space
was obtained for arbitrary $\epsilon$~\cite{omega2}, an explicit expression
for it is known only at $\epsilon\rightarrow 0$~\cite{FFK96,Blum98,DDG01}.

The NLO correction to $R$ is~\cite{FFKP00}
\[
R^{(1)}(\vec q_1, \vec q) = \frac{\omega^{(1)}(t)}{2}
\left[\frac{\epsilon \Gamma(1+2\epsilon)(\vec q^{\:2} )^{1-\epsilon}}
{2\Gamma^2(1+\epsilon)}\int\frac{d^{D-2}k}{\Gamma(1-\epsilon)\pi^{1+\epsilon}}
\frac{\ln(\vec q^{\:2} /\vec k^{\:2})}{(\vec k - \vec q_1)^2(\vec k +
\vec q_1^{\:\prime})^2}
\right.
\]
\[\left.
+ \left( \left(\frac{\vec q_1^{\:2}}{\vec q^{\:2}}\right)^\epsilon+
\left(\frac{\vec q_1^{\:\prime\:2}}{\vec q^{\:2}}\right)^\epsilon-1 \right)
\left(\frac{1}{2\epsilon}+\psi(1+2\epsilon)-\psi(1+\epsilon)+\frac{11+7\epsilon}
{2(1+2\epsilon)(3+2\epsilon)}\right) \right.
\]
\begin{equation}
\left.
-\frac{1}{2\epsilon}+\psi(1)
+\psi(1+\epsilon)-\psi(1-\epsilon)-\psi(1+2\epsilon)
\right] \;.
\label{R_1loop}
\end{equation}
The NLO corrections to the part of the kernel
related with the real particle production, ${\cal K}_r$,
can be presented~\cite{FFK00} as follows in the limit $\epsilon\rightarrow 0$:
\begin{equation}
{\cal K}_r^{(1)} = \frac{\bar g^4}{\pi^{1+\epsilon}
\Gamma(1-\epsilon)} \: \biggl( {\cal K}_1 + {\cal K}_2 + {\cal K}_3 \biggr)\;,
\;\;\;\;\;\;\;\;\;\;
\bar g^2 \equiv \frac{g^2 N_c \Gamma(1-\epsilon)}{(4 \pi)^{D/2}}\;,
\label{K_1loop}
\end{equation}
with
\begin{equation}
{\cal K}_1 = -f_B(\vec q_1,\vec q_2; \vec q) \frac{(\vec k^{\:2})^\epsilon}{\epsilon}
\left[\frac{11}{3}+\left(2 \psi'(1)-\frac{67}{9}\right)\epsilon
+\left(\frac{404}{27}+7 \psi''(1)-\frac{11}{3}\psi'(1)\right)\epsilon^2\right]\;,
\label{K1}
\end{equation}
where $\vec k=\vec q_1-\vec q_2$,
\[
{\cal K}_2 = \left\{\vec q^{\:2}\left[ \frac{11}{6}\ln\left(\frac{\vec q_1^{\:2}
\vec q_2^{\:2}}{\vec q^{\:2} \vec k^{\:2}}\right)
+\frac{1}{4}\ln\left(\frac{\vec q_1^{\:2}}{\vec q^{\:2}}\right)
\ln\left(\frac{\vec q_1^{\:\prime\:2}}{\vec q^{\:2}}\right)
+\frac{1}{4}\ln\left(\frac{\vec q_2^{\:2}}{\vec q^{\:2}}\right)
\ln\left(\frac{\vec q_2^{\:\prime\:2}}{\vec q^{\:2}}\right)
+\frac{1}{4}\ln^2\left(\frac{\vec q_1^{\:2}}{\vec q_2^{\:2}}\right)\right]
\right.
\]
\begin{equation}
\left.
-\frac{\vec q_1^{\:2} \vec q_2^{\:\prime\:2}+\vec q_2^{\:2}
\vec q_1^{\:\prime\:2}}{2 \vec k^{\:2}}\ln^2\left(\frac{\vec q_1^{\:2}}{\vec q_2^{\:2}}
\right) +\frac{\vec q_1^{\:2} \vec q_2^{\:\prime\:2}-\vec q_2^{\:2}
\vec q_1^{\:\prime\:2}}{\vec k^{\:2}}\ln\left(\frac{\vec q_1^{\:2}}{\vec q_2^{\:2}}
\right)\left(\frac{11}{6}-\frac{1}{4}\ln\left(\frac{\vec q_1^{\:2}
\vec q_2^{\:2}}{\vec k^4}\right)\right)\right\}+\biggl\{\vec q_i\leftrightarrow
\vec q_i^{\:\prime}\biggr\}\;,
\label{K2}
\end{equation}

\[
{\cal K}_3 = \left\{\frac{1}{2}\left[\vec q^{\:2}(\vec k^{\:2}-\vec q_1^{\:2}
-\vec q_2^{\:2})+2 \vec q_1^{\:2} \vec q_2^{\:2}
- \vec q_1^{\:2} \vec q_2^{\:\prime\:2}
- \vec q_2^{\:2} \vec q_1^{\:\prime\:2}
+\frac{\vec q_1^{\:2} \vec q_2^{\:\prime\:2}-\vec q_2^{\:2}\vec q_1^{\:\prime\:2}}
{\vec k^{\:2}}(\vec q_1^{\:2}-\vec q_2^{\:2})\right]\right.
\]
\begin{equation}
\left.
\times\int_0^1\frac{dx}{(\vec q_1(1-x)+\vec q_2 x)^2}\ln\left(\frac{\vec q_1^{\:2}
(1-x)+\vec q_2^{\:2} x}{\vec k^{\:2} x(1-x)}\right)\right\}
+\biggl\{\vec q_i \leftrightarrow \vec q_i^{\:\prime}\biggr\}\;.
\label{K3}
\end{equation}

Using the result of~\cite{FFKPg} for the integral in~(\ref{R_1loop}) in the
limit $\epsilon \rightarrow 0$,
\begin{equation}
(\vec q^{\:2})^{1-\epsilon} \int\frac{d^{D-2}k}{\Gamma(1-\epsilon)\pi^{1+\epsilon}}
\frac{\ln(\vec q^{\:2} /\vec k^{\:2})}{(\vec k - \vec q_1)^2(\vec k
+ \vec q_1^{\:\prime})^2}
= -\frac{1}{\epsilon}\ln\left(\frac{\vec q_1^{\:2} \vec q_1^{\:\prime\:2}}
{(\vec q^{\:2})^2}\right)
-\frac{1}{2} \ln^2\left(\frac{\vec q_1^{\:2}}{\vec q_1^{\:\prime\:2}}\right)
+O(\epsilon)\;,
\label{int_I5}
\end{equation}
Eqs.~(\ref{omega_2}), (\ref{R_1loop}), (\ref{K_1loop}), and taking into
account the properties of the kernel
\begin{equation}
f_B(\vec q_1,\vec 0; \vec q)=f_B(\vec q_1,\vec q; \vec q)={\cal K}_2(\vec q_1,
\vec 0; \vec q)=
{\cal K}_2(\vec q_1,\vec q; \vec q)={\cal K}_3(\vec q_1,\vec 0;\vec q)
={\cal K}_3(\vec q_1,\vec q; \vec q)=0~,
\label{gauge}
\end{equation}
we can present in this limit the bootstrap condition~(\ref{bootstrap}) in the
following form:
\[
\frac{(\vec q^{\:2})^{-2\epsilon}}{\pi^{1+\epsilon}\Gamma(1-\epsilon)}
\int\frac{d^{D-2}q_2}{\vec q_2^{\:2} \vec q_2^{\:\prime\:2}}
\left[{\cal K}_1+{\cal K}_2+{\cal K}_3 +f_B\left(-\frac{11}{6}
\ln\left(\frac{\vec q_2^{\:2} \vec q_2^{\:\prime\:2}}{\vec q_1^{\:2}
\vec q_1^{\:\prime\:2}}\right)+\frac{1}{2}\ln\left(\frac{\vec q_1^{\:2}}
{\vec q^{\:2}}\right)\ln\left(\frac{ \vec q_1^{\:\prime\:2}}{\vec q^{\:2}}
\right)\right.\right.
\]
\[
\left.\left.
-\frac{1}{2}\ln\left(\frac{\vec q_2^{\:2}}{\vec q^{\:2}}\right)
\ln\left(\frac{ \vec q_2^{\:\prime\:2}}{\vec q^{\:2}}\right)\right)\right]
=-\frac{1}{\epsilon^2}
\left[\frac{11}{3}+\left(2 \psi'(1)-\frac{67}{9}\right)
\epsilon+\left(\frac{404}{27}+\psi''(1)-\frac{22}{3}\psi'(1)\right)\epsilon^2\right]
\]
\begin{equation}
-\frac{2}{\epsilon}
\left[\frac{11}{3}+\left(2 \psi'(1)-\frac{67}{9}\right)\epsilon \right]
\ln\left(\frac{\vec q_1^{\:2} \vec q_1^{\:\prime\:2}}{(\vec q^{\:2})^2}\right)
-\frac{22}{3}\left(\ln^2\left(\frac{\vec q_1^{\:2}}{\vec q^{\:2}}\right)+
\ln^2\left(\frac{\vec q_1^{\:\prime\:2}}{\vec q^{\:2}}\right) \right)
\;.
\label{bootstrap'}
\end{equation}
It is easy to see that due to~(\ref{gauge}) the only divergent integral here
is that of
${\cal K}_1$. The divergent part of this integral should cancel the terms
of order $1/\epsilon$ and $1/\epsilon^2$ which appear explicitly in the R.H.S.
of~(\ref{bootstrap'}). Indeed, we have (for details of the calculation see
Appendix~A) in the limit $\epsilon \rightarrow 0$
\[
\frac{(\vec q^{\:2})^{-2\epsilon}}{\pi^{1+\epsilon}\Gamma(1-\epsilon)}
\int\frac{d^{D-2}q_2}{\vec q_2^{\:2} \vec q_2^{\:\prime\:2}} f_B
\frac{(\vec k^{\:2})^\epsilon}{\epsilon}
 =\frac{1}{\epsilon^2}
\left[1+2\epsilon \ln\left(\frac{\vec q_1^{\:2} \vec q_1^{\:\prime\:2}}
{(\vec q^{\:2})^2}\right)\right.
\]
\begin{equation}
\left.+\epsilon^2\left(2\ln^2\left(\frac{\vec q_1^{\:2}}{\vec q^{\:2}}\right)+
2\ln^2\left(\frac{\vec q_1^{\:\prime\; 2}}{\vec q^{\:2}}\right)
+\ln\left(\frac{\vec q_1^{\:2}}{\vec q^{\:2}}\right)
\ln\left(\frac{ \vec q_1^{\:\prime\:2}}{\vec q^{\:2}}\right)
-\psi^{\prime}(1)\right)\right]
\;,
\label{int_K1}
\end{equation}
and therefore~(\ref{bootstrap'}) takes the form
\[
\frac{(\vec q^{\:2})^{-\epsilon}}{\pi^{1+\epsilon}\Gamma(1-\epsilon)}
\int\frac{d^{D-2}q_2}{\vec q_2^{\:2} \vec q_2^{\:\prime\:2}}
\left[{\cal K}_2+{\cal K}_3 +f_B\left(-\frac{11}{6}
\ln\left(\frac{\vec q_2^{\:2} \vec q_2^{\:\prime\:2}}{\vec q_1^{\:2}
\vec q_1^{\:\prime\:2}}\right)+\frac{1}{2}\ln\left(\frac{\vec q_1^{\:2}}
{\vec q^{\:2}}\right)\ln\left(\frac{ \vec q_1^{\:\prime\:2}}
{\vec q^{\:2}}\right)\right.\right.
\]
\begin{equation}
\left.\left.
-\frac{1}{2}\ln\left(\frac{\vec q_2^{\:2}}{\vec q^{\:2}}\right)
\ln\left(\frac{ \vec q_2^{\:\prime\:2}}{\vec q^{\:2}}\right)\right)\right]
= 6 \psi''(1)
+\frac{11}{3}\ln\left(\frac{\vec q_1^{\:2}}{\vec q^{\:2}}\right)
\ln\left(\frac{\vec q_1^{\:\prime\:2}}{\vec q^{\:2}}\right)
\;.
\label{bootstrap''}
\end{equation}
Although due to~(\ref{gauge}) the integrals in the L.H.S. of~(\ref{bootstrap''})
are finite at $\epsilon\rightarrow 0$,
we retain the dimensional regularization, since separate terms in them are
still divergent.
Using known integrals
\begin{eqnarray}
I_1(\vec q) &=&
\frac{(\vec q^{\:2})^{1-\epsilon}}{\pi^{1+\epsilon}\Gamma(1-\epsilon)}
\int d^{D-2}q_2 \;\frac{1}{\vec q_2^{\:2} (\vec q_2-\vec q)^2}
= \frac{\Gamma^2(\epsilon)}{\Gamma(2\epsilon)} =
\frac{2}{\epsilon}+O(\epsilon)\;, \label{int_I1} \\
I_2(\vec q) &=&
\frac{(\vec q^{\:2})^{1-\epsilon}}{\pi^{1+\epsilon}\Gamma(1-\epsilon)}
\int d^{D-2}q_2 \;\frac{\ln(\vec q_2^{\:2}/\vec q^{\:2})}
{\vec q_2^{\:2} (\vec q_2-\vec q)^2}
\nonumber \\
&=& \frac{\Gamma^2(\epsilon)}{\Gamma(2\epsilon)}
\biggl[\psi(\epsilon)-\psi(2\epsilon)+\psi(1)-\psi(1-\epsilon)\biggr]
= -\frac{1}{\epsilon^2}+\psi'(1)+O(\epsilon)\;, \label{int_I2} \\
I_3(\vec q) &=&
\frac{(\vec q^{\:2})^{1-\epsilon}}{\pi^{1+\epsilon}\Gamma(1-\epsilon)}
\int d^{D-2}q_2 \;\frac{\ln^2(\vec q_2^{\:2}/\vec q^{\:2})}
{\vec q_2^{\:2} (\vec q_2-\vec q)^2}
\nonumber \\
&=& \frac{\Gamma^2(\epsilon)}{\Gamma(2\epsilon)}
\biggl[\biggl(\psi(\epsilon)-\psi(2\epsilon)+\psi(1)-\psi(1-\epsilon)\biggr)^2
+\psi'(\epsilon)-\psi'(2\epsilon)
\nonumber \\
&-& \psi'(1)+\psi'(1-\epsilon)\biggr]
= \frac{2}{\epsilon^3} - \frac{2\psi'(1)}{\epsilon}-2 \psi''(1)+O(\epsilon)\;,
\label{int_I3} \\
I_4(\vec q) &=&
\frac{(\vec q^{\:2})^{1-\epsilon}}{\pi^{1+\epsilon}\Gamma(1-\epsilon)}
\int d^{D-2}q_2 \;\frac{\ln(\vec q_2^{\:2}/\vec q^{\:2})
\ln((\vec q_2-\vec q)^2/\vec q^{\:2})}{\vec q_2^{\:2} (\vec q_2-\vec q)^2}
\nonumber \\
&=& \frac{\Gamma^2(\epsilon)}{\Gamma(2\epsilon)}
\biggl[\biggl(\psi(\epsilon)-\psi(2\epsilon)+\psi(1)-\psi(1-\epsilon)\biggr)^2
+ \psi'(1-\epsilon)-\psi'(2\epsilon)\biggr]
\nonumber \\
&=& -2 \psi''(1)+O(\epsilon)\;, \label{int_I4}
\end{eqnarray}
which can be easily calculated with the help of the
generalized Feynman parametrization (see Eq.~(\ref{feynman}) in Appendix~A), 
and~(\ref{int_I5}), we obtain the bootstrap condition in the form:
\[
\left[
\left\{-3 \psi''(1)
+\frac{1}{24} \ln^3\left(\frac{\vec q_1^{\:2}}{\vec q^{\:2}}\right)
- \frac{1}{4}\ln^2\left(\frac{\vec q_1^{\:2}}{\vec q^{\:2}}\right)
\ln\left(\frac{\vec q_1^{\:\prime\:2}}{\vec q^{\:2}}\right)
-\frac{1}{2}I(\vec q_1^{\:\prime\:2},\vec q^{\:2};\vec q_1^{\:2} )\right.\right.
\]
\begin{equation}
\left.\left.
+\frac{1}{2}I(\vec q^{\:2},\vec q_1^{\:2};\vec q_1^{\:\prime\:2} )
-\frac{3}{4}J(\vec q_1^{\:2},\vec q^{\:2}; \vec q_1^{\:\prime\:2} ) \right\}
+\biggl\{\vec q_1 \leftrightarrow \vec q_1^{\:\prime} \biggr\} \right]
+\frac{1}{\pi}
\int\frac{d\vec q_2}{\vec q_2^{\:2} \vec q_2^{\:\prime\:2}}\:{\cal K}_3 =0\;,
\label{bootstrap1}
\end{equation}
where
\begin{equation}
I(\vec p_1^{\:2}, \vec p_2^{\:2};(\vec p_1-\vec p_2)^2) =
\frac{(\vec p_1-\vec p_2)^2}{\pi}
\int d\vec p \;\frac{\ln(\vec p_2^{\:2}/\vec p^{\:2})\ln((\vec p_1-\vec p_2)^2
/(\vec p-\vec p_2)^2)}
{(\vec p-\vec p_1)^2(\vec p-\vec p_2)^2}
\;, \label{intI}
\end{equation}
\begin{equation}
J(\vec p_1^{\:2}, \vec p_2^{\:2};(\vec p_1-\vec p_2)^2 ) =
\frac{(\vec p_1-\vec p_2)^2}{\pi}
\int d\vec p \;\frac{\ln(\vec p_1^{\:2}/\vec p^{\:2})\ln(\vec p_2^{\:2}/
\vec p^{\:2})}{(\vec p-\vec p_1)^2(\vec p-\vec p_2)^2}
\;, \label{intJ}
\end{equation}
and ${\cal K}_3 $ is defined in~(\ref{K3}).

\section{Proof of the fulfillment of the bootstrap condition}

Although all integrals in~(\ref{bootstrap1}) can in principle be calculated
analytically, arising expressions are so long and cumbersome that we restrict
ourselves to present them in terms of one-dimensional integrals. The
derivation of such representations for the integrals $I$ and $J$ 
defined in Eqs.~(\ref{intI}) and (\ref{intJ}) is given in Appendix~A and 
for the integral with ${\cal K}_3$ in Appendix~B.

Moreover, the proof of the bootstrap relation can be simplified in the following
way. The integrals entering the bootstrap condition, as well as explicitly
presented logarithms, are functions of three variables: $q_1^2\equiv
-\vec q_1^{\:2},\;q_1^{\:\prime\:2} \equiv -\vec q_1^{\:\prime\:2}$
and $q^2\equiv -\vec q^{\:2}$~. Let us fix $q_1^{\:2}\leq 0$,
$q_1^{\:\prime\:2}\leq 0$ and consider the dependence from $q^2$. All integrals
entering the bootstrap condition~(\ref{bootstrap1}), as well as explicit logarithms
presented there, are real analytical functions of $q^2$ with the cut
$0 \leq q^2 < \infty$. They are determined, up to a constant (from the point
of view of dependence from $ q^2 $; but it still can be a function
of $q_1^2$ and $q_1^{\:\prime\:2}$), by their discontinuities on the cut.
The calculation of the discontinuities on the cut is a much easier problem than
the calculation of the whole functions; therefore, let us prove the bootstrap
relation for the discontinuities. If we manage to do it, the only thing
which will remain to be done is to prove the bootstrap at some fixed point in the
$q^2$ plane. Taking this point at $\infty$, we again simplify the calculations
drastically.

Let start with the discontinuities. For real functions they are equal to
their imaginary parts on the upper edge of the cut multiplied by $2i$.
The values of the functions on the upper edge of the cut are given by the
expressions obtained for $q^2 = - \vec q^{\:2} \leq 0$ with the substitution
$\vec q^{\:2}\rightarrow -q^2-i0$. Therefore for the logarithmic functions
explicitly presented in~(\ref{bootstrap1}) the calculation of the
discontinuities is trivial. For the integrals $I$ is easy to obtain from the
representations given in Appendix~A:
\[
\frac{1}{\pi} \Im I( \vec q_1^{\:2},-q^2-i0; \vec q_1^{\:\prime\:2}) =
2 \psi'(1)+\frac{1}{2}\ln \left(\frac{1-a}{a}\right)\ln (a(1-a))
+{\mathrm{Li}}_2 (a)-{\mathrm{Li}}_2 (1-a)
\]
\begin{equation}
+\int_0^a\frac{dx}{x(1-x)}\ln \left(1-\frac{x}{a}+\frac{x(1-x)
\vec q_1^{\:\prime\:2}}{q^2}\right)~,
\label{i1}
\end{equation}
\[
\frac{1}{\pi} \Im I(-q^2-i0,\vec q_1^{\:2};\vec q_1^{\:\prime\:2} )=
\psi'(1)+\ln \left(\frac{1-a}{a}\right)\ln(1-a)+{\mathrm{Li}}_2 (a)
-{\mathrm{Li}}_2 (1-a)
\]
\begin{equation}
+\int_0^a\frac{dx}{x(1-x)}\ln \left(1-\frac{x}{a}+\frac{x(1-x)
\vec q_1^{\:\prime\:2}}{q^2}\right)~,\label{i2}
\end{equation}
where
\begin{equation}
a=\frac{q^2}{\vec q_1^{\:2}+q^2}~,\;\;\;\;\;\;\;\;\;\;
{\mathrm{Li}}_2 (z) =-\int_0^z\frac{dx}{x}\ln (1-x)~.
\end{equation}
The bootstrap relation contains these integrals in the combination
for which the imaginary part is quite simple:
\begin{equation}
-\frac{1}{2}\frac{1}{\pi} \Im I(\vec q_1^{\:2},-q^2-i0; \vec q_1^{\:\prime\:2})
+\frac{1}{2}\frac{1}{\pi} \Im I(-q^2-i0,\vec q_1^{\:2}; \vec q_1^{\:\prime\:2})
=-\frac{\psi'(1)}{2}+\frac{1}{4}\ln^2\left(\frac{\vec q_1^{\:2}}{ q^2}\right)~.
\end{equation}
For the integral $J$ the representation of Appendix~A gives:
\[
\frac{1}{\pi} \Im J(\vec q_1^{\:2},-q^2-i0; \vec q_1^{\:\prime\:2} )=
2\psi'(1)+\frac{1}{2} \ln \left(\frac{1-a}{a}\right)\ln (a(1-a))
+{\mathrm{Li}}_2 (a)-{\mathrm{Li}}_2 (1-a)
\]
\begin{equation}
-2{\mathrm{Li}}_2 \left(\frac{-a}{1-a}\right)+2\int_0^a\frac{dx}{x(1-x)}
\ln \left(1-\frac{x}{a}+\frac{x(1-x)\vec q_1^{\:\prime\:2}}{q^2}\right)~.
\label{i3}
\end{equation}
The integral terms in~(\ref{i3}) give dilogarithms with rather complicated
arguments:
\[
\int_0^a\frac{dx}{x(1-x)}\ln \left(1-\frac{x}{a}+\frac{x(1-x)
\vec q_1^{\:\prime\:2}}{q^2}\right)
=- 2{\mathrm{Li}}_2 \left(\frac{\kappa^+-\vec q_1^{\:\prime\:2}}{q^2
+\vec q_1^{\:2}}\right)
-2{\mathrm{Li}}_2 \left(\frac{\kappa^--\vec q_1^{\:\prime\:2}}{q^2
+\vec q_1^{\:2}}\right)
\]
\[
-{\mathrm{Li}}_2 \left(\frac{\vec q_1^{\:\prime \:2}}{\kappa^+}\right)
+{\mathrm{Li}}_2 \left(\frac{\vec q_1^{\:2}}{\kappa^+}\right)
+\ln\left(\frac{\kappa^+-\vec q_1^{\:\prime \:2}}{q^2}\right)
\ln\left(\frac{\kappa^+}{\kappa^-}\right)
+\frac{1}{2}\ln^2\left(\frac{\kappa^+}{q^2}\right)
\]
\begin{equation}
- \ln\left(\frac{\kappa^+}{q^2}\right)
\ln\left(\frac{\vec q_1^{\:2}\vec q_1^{\:\prime\:2}}{(q^2)^2}\right)
-2\ln^2\left(\frac{q^2+\vec q_1^{\:2}}{ q^2}\right)
+2\ln\left(\frac{q^2+\vec q_1^{\:2}}{ q^2}\right)
\ln\left(\frac{\vec q_1^{\:2}\vec q_1^{\:\prime\:2}}{(q^2)^2}\right)
-\frac{1}{2}\ln^2\left(\frac{\vec q_1^{\:2}}{ q^2}\right)
~,
\end{equation}
where
\begin{equation}
\kappa^{\pm}=\frac{1}{2}\left(q^2+\vec q_1^{\:2}+\vec q_1^{\:\prime\:2}
\pm \sqrt{(q^2+\vec q_1^{\:2}+\vec q_1^{\:\prime\:2})^2-4\vec q_1^{\:2}\vec q_1^{\:\prime\:2}}\right)~.
\end{equation}
But their sum with the terms $(\vec q_1 \leftrightarrow
\vec q_1^{\:\prime})$, with account of~(\ref{identity}), drastically simplifies:
\[
\int_0^a\frac{dx}{x(1-x)}\ln \left(1-\frac{x}{a}+\frac{x(1-x)
\vec q_1^{\:\prime\:2}}{q^2}\right)
+ (\vec q_1^{\:2}\leftrightarrow \vec q_1^{\:\prime\:2})
=2{\mathrm{Li}}_2 \left(1-a \right)
\]
\begin{equation}
-\frac{1}{2}\ln\left(\frac{\kappa^-}{q^2}\right)\ln\left(\frac{\kappa^+}
{q^2}\right)-2\psi'(1)
+\ln^2(a)
-\frac{1}{2}\ln^2\left(\frac{1-a}{a}\right)+(\vec q_1^{\:2}\leftrightarrow
\vec q_1^{\:\prime\:2})~.
\label{i(1+2)}
\end{equation}
Using this relation and the identity
\begin{equation}
{\mathrm{Li}}_2 (a)+3{\mathrm{Li}}_2 (1-a)
-2{\mathrm{Li}}_2 \left(\frac{-a}{1-a}\right)=3\psi'(1)
+\ln\left(\frac{1-a}{a}\right)
\ln\left(\frac{1-a}{a^2}\right)-2\ln^2(a)~,
\end{equation}
which one can easily check comparing both sides of the identity at
$a=0$ and their derivatives with respect to $a$, we obtain from~(\ref{i3})
\[
\frac{1}{\pi} \Im J(\vec q_1^{\:2},-q^2-i0; \vec q_1^{\:\prime\:2} )+
(\vec q_1^{\:2}\leftrightarrow \vec q_1^{\:\prime\:2})
\]
\begin{equation}
= -\ln\left(\frac{\kappa^-}{q^2}\right)\ln\left(\frac{\kappa^+}{q^2}\right)
+\frac{1}{2}\ln^2\left(\frac{\vec q_1^{\:2}}{q^2}\right)+
\psi'(1)+(\vec q_1^{\:2}\leftrightarrow \vec q_1^{\:\prime\:2})~.
\label{i3'}
\end{equation}

The discontinuity of the term with ${\cal K}_3$
in~(\ref{bootstrap1}) can be calculated using the representations
obtained in Appendix~B. But it seems that the easiest way to find
it is to rewrite the integral with ${\cal K}_3$
in~(\ref{bootstrap1}) in the Minkowski space and to take the
discontinuity according to Cutkosky rules. This way is described
in Appendix~C. In any case, the result is
\begin{equation}
\frac{\Im A}{\pi}=-\frac{3}{2}\ln\left(\frac{\kappa^-}{q^2}\right)
\ln\left(\frac{\kappa^+}{q^2}\right)
+\frac{1}{4}\ln^2\left(\frac{\vec q_1^{\:2}\vec q_1^{\:\prime\:2}}{(q^2)^2}
\right)+\frac{1}{2}
\ln\left(\frac{\vec q_1^{\:2}}{q^2}\right)
\ln\left(\frac{\vec q_1^{\:\prime\:2}}{q^2}\right)~.
\label{ImA}
\end{equation}
Here $A$ is the analytic continuation (see~(\ref{A})) of the term with
${\cal K}_3$ in the bootstrap relation~(\ref{bootstrap1}).

The relation imposed by~(\ref{bootstrap1}) on the imaginary parts is:
\[
\left\{\left[-\frac{1}{8}\ln^2\left(\frac{\vec q_1^{\:2}}{q^2}\right)
+\frac{5\pi^2}{24}-\frac{1}{2}
\ln\left(\frac{\vec q_1^{\:2}}{q^2}\right)
\ln \left(\frac{\vec q_1^{\:\prime\:2}}{q^2}\right)-\frac{1}{2}\frac{1}{\pi}
\Im I( \vec q_1^{\:2},-q^2-i0; \vec q_1^{\:\prime\:2})
\right.\right.
\]
\begin{equation}
\left.\left.
+\frac{1}{2}\frac{1}{\pi} \Im I(-q^2-i0,\vec q_1^{\:2};
\vec q_1^{\:\prime\:2} )-\frac{3}{4}\frac{1}{\pi}
\Im J(\vec q_1^{\:2},-q^2-i0; \vec q_1^{\:\prime\:2})\right]+\biggl[\vec q_1^{\:2}
\leftrightarrow \vec q_1^{\:\prime\:2}\biggr]\right\}+\frac{\Im A}{\pi}=0.
\end{equation}
Using~(\ref{i(1+2)}), (\ref{i3'}) and (\ref{ImA}) it is easy to see that
this relation is satisfied.

Now in order to complete the proof of fulfillment of the bootstrap condition
it is sufficient to prove that it is satisfied in the limit $\vec q^{\:2}\gg
\vec q_1^{\:2}, \; \vec q^{\:2}\gg \vec q_1^{\:\prime\:2}$. From the expressions
for the integrals $I$ and $J$ obtained in Appendix~A we obtain in this limit
\begin{equation}
I( \vec q_1^{\:2},\vec q^{\:2}; \vec q_1^{\:\prime\:2})\simeq -\zeta(2)
\ln\left(\frac{\vec q^{\:2}}{\vec q_1^{\:2} }\right)+2\zeta(3)~,
\label{i1as}
\end{equation}
\begin{equation}
I( \vec q^{\:2},\vec q_1^{\:2}; \vec q_1^{\:\prime\:2})\simeq -\frac{1}{6}
\ln^3\left(\frac{\vec q^{\:2}}{\vec q_1^{\:2} }\right)-\zeta(2)
\ln\left(\frac{\vec q^{\:2}}{\vec q_1^{\:2} }\right)+2\zeta(3)~,
\label{i2as}
\end{equation}
and
\begin{equation}
J( \vec q_1^{\:2},\vec q^{\:2}; \vec q_1^{\:\prime\:2})\simeq -\frac{1}{6}
\ln^3\left(\frac{\vec q^{\:2}}{\vec q_1^{\:2} }\right)-2\zeta(2)
\ln\left(\frac{\vec q^{\:2}}{\vec q_1^{\:2} }\right)+4\zeta(3)~,
\label{i3as}
\end{equation}
where $\zeta (n)$ is the Riemann zeta-function, $\zeta (2)=\psi'(1)=\pi^2/6~,\;\;
2\zeta (3)=-\psi''(1)$.

The value of the contribution of the term with ${\cal K}_3$ in this limit is
found in Appendix~B:
\[
\frac{1}{\pi}
\int\frac{d\vec q_2}{\vec q_2^{\:2} \vec q_2^{\:\prime\:2}}\:{\cal K}_3\simeq
-\frac{1}{4} \ln \left(\frac{\vec q^{\:2}}{\vec q_1^{\:\prime\:2} }\right)
\ln\left(\frac{\vec q^{\:2}}{\vec q_1^{\:2} }\right)\left( \ln
\left(\frac{\vec q^{\:2}}{\vec q_1^{\:2} }\right)+\ln \left(\frac{\vec q^{\:2}}
{\vec q_1^{\:\prime\:2} }\right) \right)
\]
\begin{equation}
 -\frac{3\zeta(2)}{2}\left(
\ln \left(\frac{\vec q^{\:2}}{\vec q_1^{\:2} }\right)+
\ln \left(\frac{\vec q^{\:2}}{\vec q_1^{\:\prime\:2} }\right)\right)-6 \zeta(3)~.
\label{Aas}
\end{equation}
Putting the expressions~(\ref{i1as})-(\ref{Aas}) into the
equation~(\ref{bootstrap1}) one can easily see that it is satisfied in the
limit of large $\vec q^{\:2}$.

This completes the proof of the strong bootstrap condition for the NLO
color octet kernel.

\section{Discussion}

The BFKL approach to the description of high energy processes is based on the
gluon Reggeization. More precisely, in the derivation of the
representation for scattering amplitudes and of the BFKL equation, the Reggeized
form for the amplitudes with the gluon quantum numbers in the $t$-channels
was assumed. This assumption was proved in the leading logarithmic
approximation~\cite{BLF}, but still remains a hypothesis beyond this
approximation. Now a lot of results in the BFKL approach are obtained in the
next-to-leading order for the kernel of the BFKL
equation~\cite{FL98,CC98,FF98,FFP99,FG} and the impact
factors~\cite{FFKPg,FFKPq,FM99,FIK1,BGQ,FIK2,BGK}. These results are widely
discussed in the literature. Therefore it is very desirable to prove the
hypothesis of the gluon Reggeization in the NLO or, at least, to check it as
carefully as possible.

The self-consistency of the hypothesis demands the fulfillment of bootstrap
conditions arising from the requirement that amplitudes obtained in the
BFKL approach by use of the BFKL equation in the antisymmetric color octet
state must have the Reggeized form which was assumed in the derivation. Now
the fulfillment of the NLO bootstrap conditions for the elastic scattering
amplitudes~\cite{FF98} is proved~\cite{FFP99,FFKPg,FFKPq,FFK00}.

The bootstrap conditions must be satisfied for all amplitudes involved into
the derivation of the BFKL equation, i.e. for the amplitudes of production of
any number of particles in the multi-Regge and quasi multi-Regge kinematics.
The so called strong bootstrap conditions suggested without derivation 
in~\cite{B99,BV00} appear among them~\cite{tbp}. Although the conditions for 
the elastic amplitudes are already very restrictive, so that their fulfillment 
can convince about the correctness of the hypothesis of Reggeization, the 
conditions for inelastic amplitudes are even stronger.

In this paper we have completed the proof of the strong bootstrap conditions.
The bootstrap condition proved to be satisfied is so
strong that without the Reggeization its fulfillment would seem a miracle.
Although the Reggeized form for inelastic amplitudes implies another set of 
conditions~\cite{tbp}, which are not yet proved, the fulfillment of the strong 
bootstrap heavily supports the conclusion that also these conditions 
are satisfied. Moreover, the fulfillment of these conditions has to give 
the possibility to prove the hypothesis of gluon Reggeization in the NLO.

\vskip 1.5cm \underline {Acknowledgment}: V.S.F. thanks the
Alexander von Humboldt foundation for the research award, the
Universit\"at Hamburg and DESY for their warm hospitality while a
part of this work was done. F. Caporale (Universit\`a della Calabria 
and Istituto Nazionale di Fisica Nucleare, Gruppo collegato di Cosenza)
participated to the early stages of the present work.

\appendix

\section{Appendix}

In this Appendix we give some details about the calculation of the
integral in the L.H.S. of Eq.~(\ref{int_K1}) and of the integrals
$I(\vec p_1^{\:2}, \vec p_2^{\:2};(\vec p_1-\vec p_2)^2)$ and
$J(\vec p_1^{\:2}, \vec p_2^{\:2};(\vec p_1-\vec p_2)^2)$ defined
in~(\ref{intI}) and (\ref{intJ}), respectively. The first of these
three integrals will be calculated analytically in the
$\epsilon$-expansion. The other two will instead be presented in
terms of one-dimensional integrals, since they can not be
expressed in terms of elementary functions, and their expressions
in terms of known dilogarithms   would be long and cumbersome.

A necessary ingredient in all three cases will be the generalized Feynman
parametrization, which we recall here for the convenience:
\begin{equation}
\frac{1}{a_1^{\alpha_1}\ldots a_n^{\alpha_n}}=\frac{\Gamma(\sum_i \alpha_i)}
{\Gamma(\alpha_1)\ldots\Gamma(\alpha_n)} \int_0^1\ldots\int_0^1
\frac{dx_1\ldots dx_n\:\delta(1-\sum_i x_i)\: x_1^{\alpha_1-1}\ldots
x_n^{\alpha_n-1}}{\biggl(\sum_i a_i x_i\biggr)^{\sum_i\alpha_i}}\;.
\label{feynman}
\end{equation}

Let us start from the integral in~(\ref{int_K1}), related to the contribution
of ${\cal K}_1$ to the bootstrap condition~(\ref{bootstrap'}),
\[
I_{{\cal K}_1}\equiv
\frac{(\vec q^{\:2})^{-2\epsilon}}{\pi^{1+\epsilon}\Gamma(1-\epsilon)}
\int\frac{d^{D-2}q_2}{\vec q_2^{\:2} \vec q_2^{\:\prime\:2}} f_B \frac{(\vec
k^2)^\epsilon}{\epsilon}
\]
\begin{equation}
= \frac{(\vec q^{\:2})^{-2\epsilon}}{\pi^{1+\epsilon}\Gamma(1-\epsilon)}
\int\frac{d^{D-2}q_2}{\vec q_2^{\:2} \vec q_2^{\:\prime\:2}}
\left(\frac{\vec q_1^{\:2}\vec q_2^{\:\prime\:2}+\vec q_2^{\:2}\vec
q_1^{\:\prime\:2}}{(\vec q_1-\vec q_2)^2}-\vec q^{\:2}\right)
\frac{[(\vec q_1-\vec q_2)^2]^\epsilon}{\epsilon}\;,
\end{equation}
where the definition of $f_B$, given in~(\ref{K_Born}), and $\vec k=\vec q_1
-\vec q_2$ have been used. We need to calculate $I_{{\cal K}_1}$ with
accuracy up to order $\epsilon^0$. First of all, $I_{{\cal K}_1}$ can be split into
the sum of three contributions:
\begin{eqnarray}
I_{{\cal K}_1}&=&\frac{1}{\pi^{1+\epsilon}\Gamma(1-\epsilon)}
\frac{(\vec q^{\:2})^{-2\epsilon}\,\vec q_1^{\:2}}{\epsilon}
\int\frac{d^{D-2}q_2}{\vec q_2^{\:2}
[(\vec q_2-\vec q_1)^2]^{1-\epsilon}} \nonumber \\
&+&\frac{1}{\pi^{1+\epsilon}\Gamma(1-\epsilon)}
\frac{(\vec q^{\:2})^{-2\epsilon}\,\vec q_1^{\:\prime\:2}}{\epsilon}
\int\frac{d^{D-2}q_2}{\vec q_2^{\:\prime\:2}
[(\vec q_2-\vec q_1)^2]^{1-\epsilon}} \nonumber \\
&-&\frac{1}{\pi^{1+\epsilon}\Gamma(1-\epsilon)}
\frac{(\vec q^{\:2})^{1-2\epsilon}}{\epsilon}
\int\frac{d^{D-2}q_2}{\vec q_2^{\:2}\vec q_2^{\:\prime\:2}
[(\vec q_2-\vec q_1)^2]^{-\epsilon}} \label{IK1} \\
&\equiv& I_{{\cal K}_1}^{(a)}(\vec q_1)
+ I_{{\cal K}_1}^{(a)}(\vec q_1^{\:\prime})
- I_{{\cal K}_1}^{(b)}\;, \nonumber
\end{eqnarray}
with obvious notation.
The integral $I_{{\cal K}_1}^{(a)}(\vec q_1)$
can by easily calculated for arbitrary $\epsilon$ through Feynman parametrization
and by use of
\begin{equation}
\int\frac{d^{D-2}k}{(2\pi)^{D-1}}\frac{1}{(\vec k^{\:2} - 2 \vec k \cdot \vec p +
m^2)^\alpha}=\frac{2}{(4\pi)^{D/2}}\frac{\Gamma(\alpha+1-D/2)}{\Gamma(\alpha)}
\frac{1}{(m^2-\vec p^{\:2})^{\alpha+1-D/2}}\;,
\label{basic}
\end{equation}
giving
\begin{equation}
I_{{\cal K}_1}^{(a)}(\vec q_1)
= \frac{1}{\epsilon}\frac{\Gamma(1-2\epsilon)}{\Gamma^2(1-\epsilon)}
\frac{\Gamma(\epsilon)\Gamma(2\epsilon)}{\Gamma(3\epsilon)}
\left(\frac{\vec q_1^{\:2}}{\vec q^{\:2}}\right)^{2\epsilon}\;.
\label{IK1_a}
\end{equation}
The integral in $I_{{\cal K}_1}^{(b)}$ is less trivial than the
previous one, since it contains three factors in the denominator.
Moreover, in the expression for $I_{{\cal K}_1}^{(b)}$ the
parameter $\epsilon$ appears both explicitly (in the integrand and
in the pre-factors) and implicitly, through $D$, in the
integration measure over $\vec q_2$. This creates a problem: for
the convergence of the integral over $q_2$ the real part of
$\epsilon=(D-4)/2$ has to be positive, whereas for the convergence of
the integrals in the Feynman parametrization (\ref{feynman}) with
one of $\alpha$-s equal to $-\epsilon$   we have to suppose that
$\epsilon$ has a negative real part. To escape this contradiction
let us consider
\[
I_{{\cal K}_1}^{(b)}=\frac{1}{\pi^{1+\epsilon}\Gamma(1-\epsilon)}
\frac{(\vec q^{\:2})^{1-2\epsilon}}{\epsilon}
\int\frac{d^{D-2}q_2}{\vec q_2^{\:2}\vec q_2^{\:\prime\:2} [(\vec
q_2-\vec q_1)^2]^{-\epsilon_1}}~.
\]
Since this integral is an analytic function of $\epsilon_1$ in the
vicinity of  $\epsilon_1=0$, we will calculate it supposing that
the real part of $\epsilon_1$ is negative  (and keeping
$|\epsilon_1|\sim \epsilon$) and put $\epsilon_1= \epsilon$ after
this.

The first step for the calculation of this integral is to apply
the generalized Feynman parametrization~(\ref{feynman}) and to
perform the integration over $\vec q_2$ by use of~(\ref{basic});
this gives
\[
I_{{\cal K}_1}^{(b)}
=-\frac{\Gamma(1-\epsilon_1-\epsilon)}{\Gamma(1-\epsilon_1)\Gamma(1-\epsilon)}
\:\frac{\epsilon_1}{\epsilon}\:(\vec q^{\:2})^{1-2\epsilon}
\]
\[
\times \int_0^1 dx \int_0^1 dy \frac{(1-y)^{\epsilon+
\epsilon_1}\,
y^{-1-\epsilon_1}}{[y(x \vec q_1^{\:\prime\:2}+(1-x)\vec q_1^{\:2})+x(1-x)(1-y)
\vec q^{\:2}]^{1-\epsilon-\epsilon_1}}
\]
\[
= -\frac{\Gamma(1-\epsilon_1-\epsilon)}{\Gamma(1-\epsilon_1)\Gamma(1-\epsilon)}
\:\frac{\epsilon_1}{\epsilon}\:(\vec q^{\:2})^{1-2\epsilon}
\]
\[
\times \left[\int_0^1 dx \int_0^1 dy
\frac{[(1-y)^{\epsilon+\epsilon_1}-(1-y)^{-\epsilon}]\,
y^{-1-\epsilon_1}}{[y(x \vec q_1^{\:\prime\:2}+(1-x)\vec q_1^{\:2})+x(1-x)(1-y)
\vec q^{\:2}]^{1-\epsilon-\epsilon_1}}\right.
\]
\begin{equation}
\left.+\int_0^1 dx \int_0^1 dy
\frac{(1-y)^{-\epsilon}\,
y^{-1-\epsilon_1}}{[y(x \vec q_1^{\:\prime\:2}+(1-x)\vec q_1^{\:2})+x(1-x)(1-y)
\vec q^{\:2}]^{1-\epsilon-\epsilon_1}}\right]\;.
\end{equation}
The last equality was done for convenience because the first
integral in square brackets is $O(\epsilon)$ and can be neglected,
while in the second one, the integration over $y$ can be easily
done by using backwards the Feynman
parametrization~(\ref{feynman}):
\[
\int_0^1 dx \int_0^1 dy \frac{(1-y)^{-\epsilon}\,
y^{-1-\epsilon_1}}{[y(x \vec q_1^{\:\prime\:2}+(1-x)\vec q_1^{\:2})+x(1-x)(1-y)
\vec q^{\:2}]^{1-\epsilon-\epsilon_1}}
\]
\[
=-\frac{1}{\epsilon_1}\frac{\Gamma(1-\epsilon_1)\Gamma(1-\epsilon)}
{\Gamma(1-\epsilon-\epsilon_1)}
\int_0^1 dx \frac{[x \vec q_1^{\:\prime\:2}+(1-x) \vec q_1^{\:2}]^{\epsilon_1}}
{[x(1-x) \vec q^{\:2}]^{1-\epsilon}}\;.
\]
So, replacing back $\epsilon_1$ with $\epsilon$, we can write with $O(\epsilon^0)$
accuracy
\begin{equation}
I_{{\cal K}_1}^{(b)}=\frac{(\vec q^{\:2})^{1-2\epsilon}}{\epsilon}
\int_0^1 dx \frac{[x \vec q_1^{\:\prime\:2}+(1-x) \vec q_1^{\:2}]^\epsilon}
{[x(1-x) \vec q^{\:2}]^{1-\epsilon}}\;.
\end{equation}
The one-dimensional integral in the previous expression is needed up
to $O(\epsilon)$ and its calculation can be performed by splitting
the integration region into three parts, (i) $0<x<\delta$,
(ii) $\delta<x<1- \delta$ and (iii) $1-\delta<x<1$, and by calculating
the three resulting integrals in the limit $\delta \rightarrow 0$.
This can be done without any difficulty and leads finally to
\begin{equation}
I_{{\cal K}_1}^{(b)}=\frac{1}{\epsilon}\left\{\frac{1}{\epsilon}
\left[ \left(\frac{\vec q_1^{\:2}}{\vec q^{\:2}}\right)^\epsilon
+\left(\frac{\vec q_1^{\:\prime\:2}}{\vec q^{\:2}}\right)^\epsilon\right]
+\epsilon\left(\frac{1}{2}\ln^2\frac{\vec q_1^{\:2}}{\vec q_1^{\:\prime\:2}}
-2 \psi'(1)\right)\right\}+O(\epsilon)\;.
\label{IK1_b}
\end{equation}
Putting the results of~(\ref{IK1_a}) and (\ref{IK1_b}) into (\ref{IK1}),
we obtain, with $O(\epsilon^0)$ accuracy,
\[
I_{{\cal K}_1}\,=\,\frac{1}{\epsilon^2}
\left[1+2\epsilon \ln\left(\frac{\vec q_1^{\:2} \vec q_1^{\:\prime\:2}}{(\vec
q^{\:2})^2}\right)\right.
\]
\begin{equation}
\left.+\epsilon^2\left(2\ln^2\left(\frac{\vec q_1^{\:2}}{\vec q^{\:2}}\right)+
2\ln^2\left(\frac{\vec q_1^{\:\prime\:2}}{\vec q^{\:2}}\right) +\ln\left(\frac{\vec
q_1^{\:2}}{\vec q^{\:2}}\right)\ln\left(\frac{ \vec q_1^{\:\prime\:2}}
{\vec q^{\:2}}\right)-\psi^{\prime}(1)\right)\right].
\end{equation}

Let us consider now the integral $I(\vec p_1^{\:2}, \vec p_2^{\:2};
(\vec p_1-\vec p_2)^2)$ defined in~(\ref{intI}):
\begin{eqnarray}
I(\vec p_1^{\:2}, \vec p_2^{\:2};(\vec p_1-\vec p_2)^2) &=&
\frac{(\vec p_1-\vec p_2)^2}{\pi}
\int d\vec p \;\frac{\ln(\vec p_2^{\:2}/\vec p^{\:2})
\ln((\vec p_1-\vec p_2)^2/(\vec p-\vec p_2)^2)}
{(\vec p-\vec p_1)^2(\vec p-\vec p_2)^2}
\nonumber \\
&=&\frac{1}{\pi}
\int d\vec k \;\frac{\ln(\vec k_2^{\:2}/\vec k^{\:2})
\ln(1/(\vec k-\vec k_2)^2)}{(\vec k-\vec k_1)^2(\vec k-\vec k_2)^2}
\equiv \tilde I(\vec k_1^{\:2}, \vec k_2^{\:2})\;,
\label{intI'}
\end{eqnarray}
where we have performed the change of integration variable
\[
\vec p \longrightarrow |\vec p_1 -\vec p_2|\, \vec k
\]
and defined
\begin{equation}
\vec k_1 \equiv \frac{\vec p_1}{ |\vec p_1 -\vec p_2|}\;, \;\;\;\;\;\;\;\;\;\;
\vec k_2 \equiv \frac{\vec p_2}{ |\vec p_1 -\vec p_2|}\;, \;\;\;\;\;\;\;\;\;\;
\biggl( (\vec k_1-\vec k_2)^2=1\biggr)\;.
\label{replace}
\end{equation}
Although $\tilde I(\vec k_1^{\:2}, \vec k_2^{\:2})$ is convergent, we
introduce nevertheless dimensional regularization, since divergences will
appear in intermediate steps of the calculation. We will consider therefore
the integral
\begin{equation}
\tilde I(\vec k_1^{\:2}, \vec k_2^{\:2}) =
\frac{1}{\pi^{1+\epsilon}\Gamma(1-\epsilon)}
\int d^{D-2} k \;\frac{\ln(\vec k_2^{\:2}/\vec k^{\:2})
\ln(1/(\vec k-\vec k_2)^2)}{(\vec k-\vec k_1)^2(\vec k-\vec k_2)^2}\;,
\label{intItilde}
\end{equation}
and keep only terms up to $O(\epsilon^0)$.
First of all, we rewrite $\tilde I$ in the following form:
\begin{equation}
\tilde I(\vec k_1^{\:2}, \vec k_2^{\:2}) =
\frac{1}{\pi^{1+\epsilon}\Gamma(1-\epsilon)}
\frac{\partial}{\partial\alpha}\frac{\partial}{\partial\beta}
\left(\int d^{D-2} k \;\frac{(\vec k_2^{\:2})^\beta}
{(\vec k-\vec k_1)^2[(\vec k-\vec k_2)^2]^{1+\alpha}(\vec k^{\:2})^\beta}
\right)_{\alpha=0,\:\beta=0}\;.
\end{equation}
Then, after Feynman parametrization and integration over $\vec k$, we obtain
\begin{eqnarray}
\tilde I(\vec k_1^{\:2}, \vec k_2^{\:2}) &=& \frac{1}{\Gamma(1-\epsilon)}
\frac{\partial}{\partial\alpha}\frac{\partial}{\partial\beta}
\left( \frac{\Gamma(1+\alpha+\beta-\epsilon)}{\Gamma(1+\alpha)
\Gamma(1+\beta)}\beta
\int_0^1dx\int_0^1dy \frac{y^\epsilon}{1-y}\left(\frac{1-x}{D_y}\right)^\alpha
\right.\nonumber \\
&& \times \left.
\left(\frac{(1-y)\vec k_2^{\:2}}{y D_y}\right)^\beta D_y^{\epsilon-1}
\right)_{\alpha=0,\:\beta=0}
\nonumber \\
&=& \frac{1}{\Gamma(1-\epsilon)}
\frac{\partial}{\partial\beta}
\left(\frac{\Gamma(1+\beta-\epsilon)}{\Gamma(1+\beta)}\beta
\int_0^1dx\int_0^1dy \frac{y^\epsilon}{1-y}
\left(\frac{(1-y)\vec k_2^{\:2}}{y D_y}\right)^\beta D_y^{\epsilon-1}
\right. \nonumber \\
&& \times\left.\left[\psi(1+\beta-\epsilon)-\psi(1)-\ln\left(\frac{D_y}{1-x}\right)
\right]\right)_{\beta=0}\;,
\end{eqnarray}
where $D_y\equiv(1-y)(x\vec k_1^{\:2}+(1-x)\vec k_2^{\:2})+yx(1-x)$. Using
the property
\[
\int_0^1dy (1-y)^{\beta-1}f(y)=\frac{f(1)}{\beta}+\int_0^1dy (1-y)^{\beta-1}
[f(y)-f(1)]
\]
and performing the derivative with respect to $\beta$, we arrive at
\[
\tilde I(\vec k_1^{\:2}, \vec k_2^{\:2}) =
\int_0^1dx\left\{\left(\psi(1-\epsilon)-\psi(1)
+\ln\left(\frac{\vec k_2^{\:2}}{D_1}\right)\right)
\left(\psi(1-\epsilon)-\psi(1)-\ln\left(\frac{D_1}{1-x}\right)\right)
D_1^{\epsilon-1}\right.
\]
\begin{equation}
+\psi'(1-\epsilon) D_1^{\epsilon-1}
+\int_0^1\frac{dy}{1-y}\left[\biggl(\psi(1-\epsilon)-\psi(1)\biggr)
\biggl(y^\epsilon D_y^{\epsilon-1}-D_1^{\epsilon-1}\biggr)\right.
\label{intItilde'}
\end{equation}
\[
\left.\left. -\ln\left(\frac{D_y}{1-x}
\right)y^\epsilon D_y^{\epsilon-1}+\ln\left(\frac{D_1}{1-x}
\right)D_1^{\epsilon-1}\right]\right\}\;,
\]
where $D_1\equiv x(1-x)$.
The one-dimensional integral in the previous expression can be easily calculated
and gives
\[
\tilde I_1 = \frac{\Gamma^2(\epsilon)}{\Gamma(2\epsilon)}
\biggl[\biggl(\psi(1-\epsilon)-\psi(1)-\psi(\epsilon)+\psi(2\epsilon)\biggr)
\biggl(\psi(1-\epsilon)-\psi(1)-2(\psi(\epsilon)-\psi(2\epsilon))
+\ln \vec k_2^{\:2}\biggr)\biggr.
\]
\begin{equation}
\biggl.+\psi'(\epsilon)-2\psi'(2\epsilon)+\psi'(1-\epsilon)\biggr]\;.
\label{intItilde_1}
\end{equation}
For the two-dimensional integral in~(\ref{intItilde'}),
\[
\tilde I_2 = \int_0^1dx\int_0^1\frac{dy}{1-y}\left[\biggl(\psi(1-\epsilon)
-\psi(1)\biggr)
\biggl(y^\epsilon D_y^{\epsilon-1}-D_1^{\epsilon-1}\biggr)\right.
\]
\begin{equation}
\left.\left. -\ln\left(\frac{D_y}{1-x}
\right)y^\epsilon D_y^{\epsilon-1}+\ln\left(\frac{D_1}{1-x}
\right)D_1^{\epsilon-1}\right]\right\}\;,
\label{intItilde_2}
\end{equation}
it is convenient to separate the integration region over $x$ into three
parts, (i) $0<x<\delta$, (ii) $\delta<x<1-\delta$, (iii) $1-\delta<x<1$,
and to calculate the three corresponding integrals in the limit $\delta
\rightarrow 0$. In the integration region (i) it is possible to make the
approximations $D_y\simeq (1-y)\vec k_2^{\:2}+x$ and $D_1\simeq x$ and a
straightforward calculation leads to the following result, with accuracy
up to $O(\epsilon^0)$:
\begin{equation}
\tilde I_2^{(i)} \simeq -
\frac{2}{\epsilon^3}-\frac{1}{3}\ln^3\delta-\psi'(1)\left(\ln\delta
-\frac{1}{\epsilon}\right)+\ln\vec k_2^{\:2}\left(-\frac{1}{\epsilon^2}
+\frac{1}{2}\ln^2\delta+\psi'(1)\right)\;.
\label{intItilde_2i}
\end{equation}
Similarly, in the integration region (iii) it is possible to make the
approximations $D_y\simeq (1-y)\vec k_2^{\:2}+1-x$ and $D_1\simeq 1-x$ and
to obtain, with accuracy up to $O(\epsilon^0)$:
\begin{equation}
\tilde I_2^{(iii)} \simeq \psi'(1)\ln\left(\frac{\vec
k_1^{\:2}}{\delta}\right)+\frac{1}{2}\psi''(1) \;.
\end{equation}
Finally, in the integration region (ii) $\epsilon$ can be put equal to zero,
since there are no divergences, and we obtain
\[
\tilde I_2^{(ii)} \simeq
\ln\delta\left[\frac{1}{2}\ln^2\left(\frac{\vec k_1^{\:2}}{\delta}\right)
+\frac{1}{2}\ln^2\left(\frac{\vec k_2^{\:2}}{\delta}\right)+2 \psi'(1)\right]
-\frac{1}{2}\int_\delta^{1-\delta}\frac{dx}{D_1}\left[\ln^2x
-\ln^2\left(\frac{D_0}{1-x}\right)\right]
\]
\[
+\int_\delta^{1-\delta}dx \ln\left(\frac{x}{1-x}\right)
\ln\left(\frac{D_1}{D_0}\right)\left[\frac{1-2x}{D_1}
+\frac{\vec k_1^{\:2}-\vec k_2^{\:2}-(1-2x)}{D_1-D_0}\right]
\]
\[
= \frac{1}{3}\ln^3\delta-\frac{1}{2}\ln^2\delta\ln\vec k_2^{\:2}
+2\psi'(1)\ln\delta+2 \psi''(1)
\]
\[
+\frac{1}{2}\int_0^1\frac{dx}{x}\ln^2\left(\frac{D_0}{\vec k_2^{\:2}}\right)
+\frac{1}{2}\int_0^1\frac{dx}{1-x}\ln^2\left(\frac{D_0}{\vec k_1^{\:2}}\right)
+\ln \vec k_2^{\:2}\int_0^1\frac{dx}{x}\ln\left(\frac{D_0}{\vec k_2^{\:2}}\right)
+\ln \vec k_1^{\:2}\int_0^1\frac{dx}{1-x}\ln\left(\frac{D_0}{\vec k_1^{\:2}}\right)
\]
\[
+ \int_0^1\frac{dx}{1-x}\ln x \ln D_0
- \int_0^1\frac{dx}{x}\ln x \ln\left(\frac{D_0}{\vec k_2^{\:2}}\right)
- 2 \int_0^1\frac{dx}{1-x}\ln(1-x) \ln\left(\frac{D_0}{\vec k_1^{\:2}}\right)
\]
\begin{equation}
- \int_0^1 dx \ln\left(\frac{x}{1-x}\right)\ln \left(\frac{D_1}{D_0}\right)
\frac{\vec k_1^{\:2}-\vec k_2^{\:2}-(1-2x)}{D_0-D_1}\;,
\label{intItilde_2ii}
\end{equation}
where $D_0\equiv x \vec k_1^{\:2} +(1-x) \vec k_2^{\:2}$ and we have used
\[
\int_0^1\frac{dx}{x}\ln(1-x)=-{\mathrm Li}_2(1)=-\psi'(1)\;,\;\;\;\;\;\;\;\;\;\;
\int_0^1\frac{dx}{x}\ln^2(1-x)=2{\mathrm Li}_3(1)=-\psi''(1)\;.
\]
Summing up the results in~(\ref{intItilde_1}) and
(\ref{intItilde_2i})-(\ref{intItilde_2ii}) to obtain $\tilde I(\vec k_1^{\:2},
\vec k_2^{\:2})$ with $O(\epsilon^0)$ accuracy, it is easy to check that all
$\delta$-dependent terms disappear and that divergences cancel each other.
The final result is the following:
\begin{equation}
I(\vec p_1^{\:2},\vec p_2^{\:2};(\vec p_1-\vec p_2)^2)
=\tilde I (\vec k_1^{\:2},\vec k_2^{\:2})=\psi'(1)\ln\vec k_1^{\:2}
+\;\biggl[\mbox{one-dimensional integrals in~(\ref{intItilde_2ii})}\biggr]\;,
\label{intItilde_fin}
\end{equation}
with $\vec k_1$ and $\vec k_2$ defined as in~(\ref{replace}) and
$D_1\equiv x(1-x)$ and $D_0\equiv x \vec k_1^{\:2} +(1-x) \vec k_2^{\:2}$.

The third integral we should consider in this Appendix is
$J(\vec p_1^{\:2}, \vec p_2^{\:2};(\vec p_1-\vec p_2)^2)$ defined in~(\ref{intJ}).
Its calculation can be carried on following the same strategy as
for $I(\vec p_1^{\:2}, \vec p_2^{\:2};(\vec p_1-\vec p_2)^2)$. We will not present
here the calculation, but merely show the final result, which reads
\[
J(\vec p_1^{\:2},\vec p_2^{\:2};(\vec p_1-\vec p_2)^2)
= \frac{(\vec p_1-\vec p_2)^2}{\pi}
\int d\vec p \;\frac{\ln(\vec p_1^{\:2}/\vec p^{\:2})\ln(\vec p_2^{\:2}
/\vec p^{\:2})}{(\vec p-\vec p_1)^2(\vec p-\vec p_2)^2}
\]
\[
= \frac{1}{\pi}
\int d\vec k \;\frac{\ln(\vec k_1^{\:2}/\vec k^{\:2})\ln(\vec k_2^{\:2}/\vec k^{\:2})}
{(\vec k-\vec k_1)^2(\vec k-\vec k_2)^2}
=\ln \left(\frac{\vec k_1^{\:2}}{\vec k_2^{\:2}}\right)
\left[\int_0^1\frac{dx}{1-x}\ln\left(\frac{D_0}{\vec k_1^{\:2}}\right)
-\int_0^1\frac{dx}{x}\ln\left(\frac{D_0}{\vec k_2^{\:2}}\right)\right]
\]
\[
+ \int_0^1\frac{dx}{x}\ln^2\left(\frac{D_0}{\vec k_2^{\:2}}\right)
+ \int_0^1\frac{dx}{1-x}\ln^2\left(\frac{D_0}{\vec k_1^{\:2}}\right)
\]
\[
- 2\int_0^1 \frac{dx}{D_1-D_0}\ln\left(\frac{x}{1-x}\right)
\ln\left(\frac{D_1}{D_0}\right)
\left[(1-2x)-\frac{D_1(\vec k_1^{\:2}-\vec k_2^{\:2})}{D_0}\right]\;,
\]
with the same notations as the previous calculation, i.e.
with $\vec k_1$ and $\vec k_2$ defined as in~(\ref{replace}) and
$D_1\equiv x(1-x)$ and $D_0\equiv x \vec k_1^{\:2} +(1-x) \vec k_2^{\:2}$.

\section{Appendix}

It is interesting to note (and can be useful for applications) that the
integral $I$ entering the contribution ${\cal K}_3$ to the kernel
is a totally symmetric function of the variables $\vec q_1^{\:2},
\vec q_2^{\:2}$ and $\vec k^{\:2}$. It is obvious from the representation
\begin{equation}
I=\int_0^1\int_0^1\int_0^1\frac{dx_1 dx_2 dx_3\delta(1-x_1-x_2-x_3)}{(\vec q_1^{\:2}x_1+
\vec q_2^{\:2}x_2+\vec k^{\:2}x_3)(x_1x_2+x_1x_3+x_2x_3)}~,
\end{equation}
which also can be useful in applications. One can easily check that this
representation reproduces the original form, performing the integration
over $x_3$, then making the change of variables
\[
x=\frac{x_2}{x_1+x_2}~, \;\;\; z=\frac{x_1x_2}{x_3(1-x_3)+x_1x_2}
=\frac{x_1x_2}{x_1(1-x_1)+x_2(1-x_2)-x_1x_2}~;
\]
\begin{equation}
x_1=\frac{(1-x)z}{z+x(1-x)(1-z)}~, \;\;\; x_2=\frac{xz}{z+x(1-x)(1-z)}~, \;\;\;
x_3=\frac{x(1-x)(1-z)}{z+x(1-x)(1-z)}~,
\end{equation}
with the Jacobian
\begin{equation}
J\left(\frac{x_1,x_2}{x,z}\right)= \frac{zx(1-x)}{(z+x(1-x)(1-z))^3}~,
\end{equation}
and integrating over $z$.

Another useful representation is
\begin{equation}
I=\int_0^1dx \int_1^\infty \frac{dt}{t} \: \frac{1}{\vec q_1^{\:2}x(1-x)(t-1)+
\vec q_2^{\:2}(1-x)+\vec k^{\:2}x}~.
\end{equation}
Using this representation it is possible in a relatively simple way to
express the contribution with ${\cal K}_3$ in the bootstrap condition
in terms of a one-dimensional integral. Let us divide this contribution
into several pieces:
\begin{equation}
A \equiv \frac{1}{\pi}\int \frac{d^2q_2}{\vec q_2^{\:2} \vec q_2^{\:\prime\:2}}
{\cal K}_3 =
(A_1+A_2+A_3)+(\vec q_1 \leftrightarrow \vec q_1^{\:\prime})~,
\end{equation}
where
\begin{equation}
A_1=\frac{1}{2\pi}\int \frac{d^2q_2}{\vec q_2^{\:2} \vec q_2^{\:\prime\:2}}
\left[\vec q^{\:2}
(\vec k^{\:2}-\vec q_1^{\:2})+\vec q_2^{\:2}(\vec q_1^{\:2}-
 \vec q_1^{\:\prime\:2})\right]~I~,
\end{equation}
\begin{equation}
A_2=\frac{1}{2\pi}\int \frac{d^2q_2}{\vec k^{\:2} \vec q_2^{\:\prime\:2}}
\left[ \vec q_1^{\:\prime\:2}
(\vec q_2^{\:2}-\vec q_1^{\:2})+\vec k^{\:2}(\vec q_1^{\:2}-
\vec q^{\:2})\right]~I~,
\end{equation}
\begin{equation}
A_3=\frac{1}{2\pi}\int \frac{d^2q_2}{\vec q_2^{\:2} \vec k^{\:2}}\vec q_1^{\:2}
(\vec q_1^{\:2}-\vec k^{\:2}-\vec q_2^{\:2})~I~.
\end{equation}
It is easy to see, with account of
the relations $\vec k=\vec q_1 -\vec q_2$ and $\vec q_{1,2}^{\:\prime}=
\vec q -\vec q_{1,2}$~, that
\begin{equation}
A_2=A_1(\vec q_1^{\:\prime} \leftrightarrow -\vec q)~, \;\;\;\;\;
A_3=-A_1(\vec q_1^{\:\prime}=0)~,
\end{equation}
so that we need to calculate only $A_1$.

Let us write the Feynman parametrization in the form:
\[
\frac{1}{\vec q_2^{\:2} \vec q_2^{\:\prime\:2}(\vec q_1^{\:2}x(1-x)(t-1)+
\vec q_2^{\:2}(1-x)+\vec k^{\:2}x)}=2\int_0^x\frac{dz}{x} \int_0^1dv (1-v)
\]
\begin{equation}
\times \frac{1}
{(\vec q_2^{\:2}(1-v)(1-z/x)+\vec q_2^{\:\prime\:2}v+(\vec q_1^{\:2}x(1-x)(t-1)
+\vec q_2^{\:2}(1-x)+\vec k^{\:2}x)(1-v)z/x)^3}~.
\end{equation}
Then, after the integration over $\vec q_2$ (of course, with account of
the relation $\vec k=\vec q_1 -\vec q_2$), the subsequent integrations over $v$
and $t$ are quite simple and we arrive at
\[
A_1=\frac{1}{2}\int_0^1\frac{dx}{x}\int_0^x \frac{dz}{(\vec q^{\:2}-
\vec q_1^{\:2}z)(1-z)+
\vec q_1^{\:\prime\:2}z }\left[((\vec q^{\:2}+\vec q_1^{\:2}-
\vec q_1^{\:\prime\:2})x-2\vec q^{\:2})\right.
\]
\[
\left.\times \left(\frac{1}{x-z}\ln\left(\frac{1-z}{1-x}\right)
-\frac{\vec q_1^{\:2}z}{\vec q^{\:2}(1-z)-\vec q_1^{\:2}z(1-x)+
\vec q_1^{\:\prime\:2}z}\ln \left(\frac{\vec q^{\:2}(1-z)+
\vec q_1^{\:\prime\:2}z}{\vec q_1^{\:2}z(1-x)}\right)\right)\right.
\]
\begin{equation}
\left.+(\vec q^{\:2}+\vec q_1^{\:2}-
\vec q_1^{\:\prime\:2})\ln \left(\frac{\vec q^{\:2}(1-z)+
\vec q_1^{\:\prime\:2}z}{\vec q_1^{\:2}z(1-z)}\right)\right]~.
\end{equation}
The result of the further integration of the total integrand in any of the
variables $x, z $ can not be expressed in terms of elementary functions. But
we can perform the integration in such a way: integrate over $x$ all
terms except those which contain $\ln (1-x)$; these terms must be integrated
over $z$ first. The terms with the denominator $x-z$ must be regularized
(for example, we can perform the integration over the region
$z \geq 0$, $x\leq 1$, $x-z \geq \delta \rightarrow 0 $. The result is
\[
A_1=\frac{1}{2}\int_0^1 \frac{dz}{\vec q^{\:2}(1-z)+ \vec q_1^{\:\prime\:2}z
-\vec q_1^{\:2}z(1-z)}\left[\left(\frac{2\vec q^{\:2}\vec q_1^{\:2}z}
{\vec q^{\:2}(1-z)+ \vec q_1^{\:\prime\:2}z-\vec q_1^{\:2}z}\right.\right.
\]
\[
\left.\left.
+\vec q^{\:2}+\vec q_1^{\:2}- \vec q_1^{\:\prime\:2}\right)
\ln \left(\frac{\vec q^{\:2}(1-z)+ \vec q_1^{\:\prime\:2}z
-\vec q_1^{\:2}z(1-z)}{z(\vec q^{\:2}(1-z)+ \vec q_1^{\:\prime\:2}z)}\right)
\ln \left(
\frac{\vec q^{\:2}(1-z)+ \vec q_1^{\:\prime\:2}z}{\vec q_1^{\:2}z}\right)\right.
\]
\begin{equation}
\left.+ \left(\frac{2\vec q^{\:2}}{z}-\vec q^{\:2}-\vec q_1^{\:2}
+ \vec q_1^{\:\prime\:2}\right)\ln (1-z)
\ln \left(\frac{\vec q^{\:2}(1-z)+ \vec q_1^{\:\prime\:2}z
-\vec q_1^{\:2}z(1-z)}{\vec q^{\:2}(1-z)}\right)\right]~.
\end{equation}

It is rather easy to obtain from this representation exact value of $A_3$. Putting
$\vec q^{\:2}=\vec q_1^{\:2}~,\;\; \vec q_1^{\:\prime\:2}=0$ we have
\begin{equation}
A_3=-2\zeta(3)~.
\end{equation}
It is not difficult as well to calculate
the asymptotic behavior of $A_1$ and $A_2$ at $\vec q^{\:2}
\gg \vec q_1^{\:2}~,\;\; \vec q^{\:2} \gg \vec q_1^{\:\prime\:2}$~. We obtain
\begin{equation}
A_1\simeq -\frac{\zeta(2)}{2}\ln \left(\frac{\vec q_1^{\:2}}{\vec q_1^{\:\prime\:2}}
\right)~,
\end{equation}
\begin{equation}
A_2\simeq -\frac{1}{4} \ln ^2 \left(\frac{\vec q^{\:2}}{\vec q_1^{\:\prime\:2} }
\right)
\ln \left(\frac{\vec q^{\:2}}{\vec q_1^{\:2} }\right)-\frac{\zeta(2)}{2}\left(
\ln \left(\frac{\vec q^{\:2}}{\vec q_1^{\:2} }\right)
+2\ln \left(\frac{\vec q^{\:2}}{\vec q_1^{\:\prime\:2} }\right)\right)- \zeta(3)~,
\end{equation}
so that
\begin{equation}
A\simeq -\frac{1}{4} \ln \left(\frac{\vec q^{\:2}}{\vec q_1^{\:\prime\:2} }\right)
\ln\left(\frac{\vec q^{\:2}}{\vec q_1^{\:2} }\right)\left( \ln
\left(\frac{\vec q^{\:2}}{\vec q_1^{\:2} }\right)+\ln \left(\frac{\vec q^{\:2}}
{\vec q_1^{\:\prime\:2} }\right)\right) -\frac{3\zeta(2)}{2}\left(
\ln\left(\frac{\vec q^{\:2}}{\vec q_1^{\:2} }\right)+\ln
\left(\frac{\vec q^{\:2}}{\vec q_1^{\:\prime\:2} }\right)\right)-6 \zeta(3)~.
\end{equation}

For calculation of the discontinuity of $A$ at $q^2=-\vec q^{\:2}\geq 0$
the last representation of $A_3$ is not very convenient, since its analytical
properties are not simple. In fact, it is more convenient to do one step back
and to use the representation in the form of the two-dimensional integral.

\section{Appendix}

Another way of calculation of the discontinuity of $A$ at
$q^2=-\vec q^{\:2}\geq 0$ is to rewrite the integral over $q_2$ in
Minkowski space and to use the Cutkosky rules for the calculation
of the discontinuity. Let us use the representation
\begin{equation}
I=\int_0^1 dx \int_0^\infty dz \frac{1}{z-k^2x(1-x)-i0}\frac{1}{z-q_1^2(1-x)
-q_2^2x-i0}~,
\end{equation}
where $k, \; q_1, \; q_2$ are considered as vectors in the two-dimensional
Minkowski space, i.e. $k^2=-\vec k^{\:2}, \;q_1^2=-\vec q_1^{\:2},\; q_2^2=
-\vec q_2^{\:2}$. This representation can be used for arbitrary values of
$k^2, \;q_1^2,\; q_2^2$. Analogous representation can be written for
$I(q_i \leftrightarrow
q_i^{\prime})$. It permits to rewrite the integral with ${\cal K}_3$ in the
bootstrap relation in the form
\begin{equation}
A = \frac{1}{i\pi}\int \frac{d^2q_2}{(q_2^{\:2}+i0)((q-q_2)^2+i0)}{\cal K}_3~,
\label{A}
\end{equation}
where now
\begin{equation}
d^2q_2=dq_2^{(0)}dq_2^{(1)},\;\; q_2^{\:2}=(q_2^{(0)})^2-(q_2^{(1)})^2~,
\end{equation}
etc., which determines $A$ as the function of
$q_1^2,\;q_1^{\:\prime\:2}$ and $q^2$ for arbitrary values of
these variables. For $q_1^2\equiv -\vec q_1^{\:2} \leq
0,\;q_1^{\:\prime\:2} \equiv -\vec q_1^{\:\prime\:2} \leq 0$ and
$q^2\equiv -\vec q^{\:2} \leq 0$ it is just the function
entering~(\ref{bootstrap1}), that is easily seen by making the Wick
rotation of the contour of integration over $q_2^{(0)}$. We are
interested in the region $q_1^2\leq 0,\;q_1^{\:\prime\:2} \leq 0$
and $q^2 \geq 0$. According to the Cutkosky rules, the
discontinuity of $A$ related to the terms with $I$ is determined
by the two cuts, with the contributions obtained by the
substitutions:
\begin{equation}
\frac{1}{(q_2^{\:2}+i0)((q-q_2)^2+i0)}\rightarrow (-2\pi i)^2 \delta(q_2^{\:2})
 \delta((q-q_2)^2 )~
\end{equation}
and
\begin{equation}
\frac{1}{(z-q_1^2(1-x)-q_2^2x-i0)((q-q_2)^2+i0)} \rightarrow -(-2\pi i)^2
\delta(z-q_1^2(1-x)-q_2^2x)\delta((q-q_2)^2 )~.
\end{equation}
Using these rules and removing the $\delta$-functions by the integration
over $q_2$ (the most appropriate system for this is $q^{(1)}=0,\; q^2=
(q^{(0)})^2$), we obtain
\begin{equation}
\frac{\Im A}{\pi} =f_1+f_2+(q_1^2 \leftrightarrow q_1^{\:\prime \:2})~,
\end{equation}
where the contributions $f_1$ and $f_2$ come from the first and second cuts,
respectively. We obtain for them
\begin{equation}
f_1=\frac{1}{2}\int_0^1 dx \int_0^\infty dz \frac{\vec q_1^{\:2}-
\kappa^+}{(z+\kappa^+x(1-x))(z+\vec q_1^{\:2}(1-x))} +
(\kappa^+\rightarrow \kappa^-)~,
\end{equation}
where $\kappa^{\pm} $ are the values of $-k^2$ on the mass shell
$ q_2^{\:2}=0,\;(q-q_2)^2=0 $, related to the two possible solutions for
$q_2^{(1)}$ :
\begin{equation}
\kappa^{\pm}=\frac{1}{2}\left(q^2+\vec q_1^{\:2}+\vec q_1^{\:\prime\:2}
\pm \sqrt{(q^2+\vec q_1^{\:2}+\vec q_1^{\:\prime\:2})^2-4\vec q_1^{\:2}
\vec q_1^{\:\prime\:2}}\right)~,
\end{equation}
and
\[
f_2= \frac{1}{2}\int_0^1 dx \int_0^\infty \frac{dz \: \theta (q^2x
-\vec q_1^{\:2}(1-x)-z)}{z+\tilde\kappa^+x(1-x)}\left[
\frac{\kappa^+-\vec q_1^{\:2}}{z+\vec q_1^{\:2}(1-x)}\right.
\]
\begin{equation}
\left.+\frac{\kappa^+(q^2+\vec q_1^{\:2})-
\vec q_1^{\:2}\vec q_1^{\:\prime\:2}}{xq^2\tilde\kappa^+}\right]+
(\kappa^+\rightarrow \kappa^-)~,
\end{equation}
where $\tilde\kappa^{\pm} $ are given by values of $-k^2$ on the mass
shell $z-q_1^2(1-x)-q_2^2x
=0,\;(q-q_2)^2=0 $, so that
\begin{equation}
\tilde\kappa^{\pm}=\kappa^{\pm}\left(1-\frac{z+\vec q_1^{\:2}(1-x)}{q^2x}
\right)+\vec q_1^{\:\prime\:2}\frac{z+\vec q_1^{\:2}(1-x)}{q^2x}~.
\end{equation}
The integration over $z$ is quite elementary. Making subsequent integration
over $x$ one has to pay attention that the contributions $f_{1,2}$
are separately divergent at $x=1$. The cancellation of the divergent terms is done
easily and unambiguously. After that, a straightforward integration gives
\[
\frac{\Im A}{\pi} ={\mathrm{Li}}_2 \left(\frac{\kappa^+-\vec q_1^{\:\prime\:2}}
{q^2+\vec q_1^{\:2}}\right)+{\mathrm{Li}}_2 \left(\frac{\kappa^--
\vec q_1^{\:\prime\:2}}{q^2+\vec q_1^{\:2}}\right)
+{\mathrm{Li}}_2 \left(\frac{\vec q_1^{\:2}}{q^2+\vec q_1^{\:2}}\right)-\zeta(2)
+\frac{1}{8}\ln^2\left(\frac{\kappa^+}{q^2}\right)
\]
\begin{equation}
+\frac{1}{8}\ln^2\left(\frac{\kappa^-}{q^2}\right)+\frac{1}{2}
\ln\left(\frac{q^2+\vec q_1^{\:2}}{ q^2}\right)
\ln\left(\frac{ q^2(q^2+\vec q_1^{\:2})^3}{(\vec q_1^{\:2}
\vec q_1^{\:\prime\:2})^2}\right)
+ \frac{1}{4}\ln\left(\frac{\vec q_1^{\:2}}{q^2}\right)
\ln\left(\frac{\vec q_1^{\:\prime\:2}}{q^2}\right)
+(q_1^2 \leftrightarrow q_1^{\:\prime \:2})~.
\end{equation}

This expression can be considerably simplified using the identity
\[
 {\mathrm{Li}}_2 \left(\frac{\kappa^+-\vec q_1^{\:\prime\:2}}{q^2+\vec
q_1^{\:2}}\right)
+{\mathrm{Li}}_2 \left(\frac{\kappa^--\vec q_1^{\:\prime\:2}}{q^2+\vec
q_1^{\:2}}\right)
+\frac{1}{2}\ln\left(\frac{\kappa^-}{q^2}\right)\ln\left(\frac{\kappa^+}
{q^2}\right)+(q_1^2 \leftrightarrow q_1^{\:\prime \:2})
\]
\begin{equation}
=\zeta(2)-{\mathrm{Li}}_2\left(\frac{\vec q_1^{\:2}}{q^2+\vec q_1^{\:2}}\right)
-\frac{1}{2}\ln\left(\frac{q^2+\vec q_1^{\:2}}{ q^2}\right)
\ln\left(\frac{ q^2(q^2+\vec q_1^{\:2})^3}{(\vec q_1^{\:2}
\vec q_1^{\:\prime\:2})^2}\right)
+(q_1^2 \leftrightarrow q_1^{\:\prime \:2})~,
\label{identity}
\end{equation}
which is not evident, of course, but can be proved, for example, considering
both sides of the identity as the functions of $\vec q_1^{\:2}$ (evidently,
taking into account the dependence of $\kappa^\pm $ from $\vec q_1^{\:2}$) and
comparing their derivatives and their limits at $\vec q_1^{\:2}\rightarrow 0$. 
Using this identity we obtain finally
\begin{equation}
\frac{\Im A}{\pi}=-\frac{3}{2}\ln\left(\frac{\kappa^-}{q^2}\right)
\ln\left(\frac{\kappa^+}{q^2}\right)
+\frac{1}{4}\ln^2\left(\frac{\vec q_1^{\:2}\vec q_1^{\:\prime\:2}}{(q^2)^2}
\right)+\frac{1}{2}
\ln\left(\frac{\vec q_1^{\:2}}{q^2}\right)
\ln\left(\frac{\vec q_1^{\:\prime\:2}}{q^2}\right)~.
\end{equation}
Of course, the same result follows from the representation of $A$ obtained in
Appendix~B.

\end{document}